\documentclass[12pt]{article}
\usepackage{graphics}
\usepackage{graphicx}
\usepackage{color}
\usepackage[hypertex]{hyperref}
\usepackage{amsfonts}
\usepackage{eufrak}

\usepackage{amsmath}
\usepackage{amstext}
\usepackage{amsopn}
\usepackage{amsbsy}
\usepackage{amscd}
\usepackage{amsxtra}
\usepackage{amsthm}

\numberwithin{equation}{section}
\newtheorem{theorem}{Theorem}[section]
\newtheorem{lemma}[theorem]{Lemma}
\newtheorem{proposition}[theorem]{Proposition}
\newtheorem{corollary}[theorem]{Corollary}
\newtheorem{remark}[theorem]{Remark}

\newcommand{\BE}{\begin{equation}}
\newcommand{\BEN}{\begin{equation*}}
\newcommand{\EE}{\end{equation}}
\newcommand{\EEN}{\end{equation*}}
\newcommand{\BL}{\begin{lemma}}
\newcommand{\EL}{\end{lemma}}
\newcommand{\BT}{\begin{theorem}}
\newcommand{\ET}{\end{theorem}}
\newcommand{\BP}{\begin{proposition}}
\newcommand{\EP}{\end{proposition}}
\newcommand{\BC}{\begin{corollary}}
\newcommand{\EC}{\end{corollary}}
\newcommand{\BR}{\begin{remark}}
\newcommand{\ER}{\end{remark}}

\allowdisplaybreaks

\title{Ground state of nonlinear Schr\"odinger systems with saturable nonlinearity}

\author{ Tai-Chia Lin\thanks{ Department of Mathematics \& Mathematics Division, National Center for Theoretical Sciences at Taipei, National Taiwan University, Taipei, 10617, Taiwan. Email: tclin@math.ntu.edu.tw}
\ , Milivoj R. Beli\'c
\thanks{Texas A\&M University at Qatar, P.O. Box 23874, Doha,
Qatar. }\ , Milan S. Petrovi\'c
\thanks{Texas A\&M University at Qatar, P.O. Box 23874,
Doha, Qatar, Institute of Physics, P.O. Box 57, 11001 Belgrade,
Serbia}\, 
Goong Chen\thanks{Department of Mathematics, Texas A\&M University,
College Station, Texas 77843, USA. } }

\date{\today}

\begin{document}

\maketitle

\begin{abstract}
\noindent We prove the existence of ground state in a
multidimensional nonlinear Schr\"odinger model of paraxial beam
propagation in isotropic local media with saturable nonlinearity.
Such ground states exist in the form of bright counterpropagating
solitons. From the proof, a general threshold condition on the
beam coupling constant for the existence of such fundamental
solitons follows.
\end{abstract}

\section{Introduction}\label{s1}
\medskip
\noindent
The existence of solitary waves in nonlinear evolution partial differential
equations has been the major concern since the beginnings of the field
\cite{ablowitz,yuri}.
The existence of ground state or fundamental soliton in one-dimensional nonlinear
Schr\"odinger (NLS) equation was never much of a concern -- their existence and stability
followed from the inverse scattering theory \cite{ablowitz}. However, in more than one
dimension this was, and still is, an open question \cite{sulem}. In multidimensional cases
there exists no mathematically rigorous theory that
would guarantee their existence and uniqueness --
let alone stability in propagation.

In this paper we prove the existence of ground state in the form of counterpropagating
solitons in an isotropic local saturable NLS model of beam propagation \cite{LPR11}.
This model represents a physically
relevant description of soliton generation, following from the theory of photorefractive effect in
crystals that respond to light by
changing their index of refraction.

\section{The model}\label{s2}

We consider the following dimensionless NLS system:
\BE\label{eq-1.1}
    \left\{ \begin{matrix}
   i{{F}_{z}}+{{\Delta }_{\bot }}F+\Gamma {{E}_{0}}F=0,  \\
   -i{{B}_{z}}+{{\Delta }_{\bot }}B+\Gamma {{E}_{0}}B=0,  \\
   {{\partial }_{t}}{{E}_{0}}+{{E}_{0}}=-\frac{{{I}_{0}}}{1+{{I}_{0}}},  \\
   {{I}_{0}}={{\left| F \right|}^{2}}+{{\left| B \right|}^{2}},  \\
\end{matrix} \right.
\EE where $F$ and $B$ are the slowly-varying envelopes of the forward and backward propagating beams,
 $z$ ($0<z<L$) is the propagation coordinate,
 $ x=\left( {{x}_{1}},{{x}_{2}} \right) \in {{\mathbb{R}}^{2}}$ are the transverse coordinates, and ${{\Delta }_{\bot }}=\sum\limits_{j=1}^{2}{\partial _{{{x}_{j}}}^{2}}$ is the transverse Laplacian. Furthermore,
 $\Gamma$ is the beam coupling constant, $E_0$ the homogenous part of
the space charge field generated in the photorefractive crystal, and $I_0$ is the beam intensity,
expressed in terms of the background intensity $I_b$.
The time independent solution of (\ref{eq-1.1}) must satisfy
${{\partial }_{t}}{{E}_{0}}=0$,
i.e. ${{E}_{0}}=-\frac{{{I}_{0}}}{1+{{I}_{0}}}=-\frac{{{\left| F \right|}^{2}}+{{\left| B \right|}^{2}}}{1+{{\left| F \right|}^{2}}+{{\left| B \right|}^{2}}}.$
We consider such a situation. Then (\ref{eq-1.1}) becomes
\BE\label{eq-1.2}
    \left\{ \begin{matrix}
   i{{F}_{z}}+{{\Delta }_{\bot }}F-\Gamma \frac{{{\left| F \right|}^{2}}+{{\left| B \right|}^{2}}}{1+{{\left| F \right|}^{2}}+{{\left| B \right|}^{2}}}F=0,\text{  }  \\
   -i{{B}_{z}}+{{\Delta }_{\bot }}B-\Gamma \frac{{{\left| F \right|}^{2}}+{{\left| B \right|}^{2}}}{1+{{\left| F \right|}^{2}}+{{\left| B \right|}^{2}}}B=0.  \\
\end{matrix} \right.
\EE Here, we assume the following boundary and initial conditions.

\noindent Boundary conditions: $F\left( x,z \right),\text{ }B\left( x,z \right)\to 0\text{   as   }\left| x \right|\to \infty ,\text{  for  0 }z<L$.

\noindent Initial conditions: $F\left( x,0 \right)={{F}_{0}}\left( x \right),\text{ }B\left( x,L \right)={{B}_{L}}\left( x \right)\text{  are given}\text{.}$

\noindent Equations (\ref{eq-1.2}) can be written as  $i{{F}_{z}}=\frac{\delta E\left[ B,F \right]}{\delta F},\text{ }-i{{B}_{z}}=\frac{\delta E\left[ B,F \right]}{\delta B}$,
where
    \[E\left[ B,F \right]=\int_{{{\mathbb{R}}^{2}}}{\tfrac{1}{2}\left( {{\left| \nabla B \right|}^{2}}+{{\left| \nabla F \right|}^{2}} \right)+\tfrac{1}{2}\Gamma \left[ {{\left| B \right|}^{2}}+{{\left| F \right|}^{2}}-\ln \left( 1+{{\left| B \right|}^{2}}+{{\left| F \right|}^{2}} \right) \right]}\] is the energy functional. The integral (here and elsewhere) is taken across the whole transverse plane.
One basic conservation law of the system (\ref{eq-1.2}) is the power conservation, given by
\BE\label{eq-1.3}
    \int_{{{\mathbb{R}}^{2}}}{{{\left| B \right|}^{2}}+{{\left| F \right|}^{2}}}=\int_{{{\mathbb{R}}^{2}}}{{{\left| {{B}_{0}} \right|}^{2}}+{{\left| {{F}_{0}} \right|}^{2}}} \quad \text{   for   } \quad 0<z<L.
\EE
To get the standing wave profiles of the system (\ref{eq-1.2}), we set
$$
F\left(x,z\right)={{e}^{i\lambda z}}u\left( x \right),\text{  }B\left( x,z \right)={{e}^{-i\lambda z}}v\left( x \right)\,,
$$
and then the system (\ref{eq-1.2}) can be transformed into the following system:
\BE\label{eq-1.4}
           \left\{ \begin{matrix}
   {{\Delta }_{\bot }}u-\Gamma \frac{{{u}^{2}}+{{v}^{2}}}{1+{{u}^{2}}+{{v}^{2}}}u=\lambda u,\text{  }  \\
   {{\Delta }_{\bot }}v-\Gamma \frac{{{u}^{2}}+{{v}^{2}}}{1+{{u}^{2}}+{{v}^{2}}}v=\lambda v,  \\
\end{matrix} \right.
\EE
so that the condition (\ref{eq-1.3}) also can be normalized as follows:
\BE\label{eq-1.5}
       P\left[ u,v \right]=\int_{{{\mathbb{R}}^{2}}}{\left( {{u}^{2}}+{{v}^{2}} \right)}\text{=1  (normalization).}
\EE
Here, $\lambda $ is the propagation constant, which can be regarded as the chemical potential in other physical settings;
mathematically it is the Lagrange multiplier following from condition (\ref{eq-1.5}). The system of equations (\ref{eq-1.4})
has been solved numerically in \cite{milan},
but with an external lattice potential $I_g$ included instead of the uniform background intensity $I_b=1$.
Fundamental counterpropagating solitons have been obtained and a threshold condition
determined. However, these results have been numerical, without rigorous proofs to substantiate their existence.

To obtain the ground state of (\ref{eq-1.2}) rigorously, we consider the following energy minimization problem:
\BE\label{eq-1.6}
    {{\mu }_{\Gamma }}=inf\left\{ E\left[ u,v \right]:\text{ }u,v\in {{H}^{1}}\left( {{\mathbb{R}}^{2}} \right),\text{ }P\left[ u,v \right]=1 \right\},
\EE
where the energy functional now is
    \[E\left[ u,v \right]=\int_{{{\mathbb{R}}^{2}}}{\left( {{\left| \nabla u \right|}^{2}}+{{\left| \nabla v \right|}^{2}} \right)+\Gamma \left[ {{u}^{2}}+{{v}^{2}}-\ln \left( 1+{{u}^{2}}+{{v}^{2}} \right) \right]}.\]
The main issue with the problem (\ref{eq-1.6}) is whether the value ${{\mu }_{\Gamma }}$ (depending on the coupling constant $\Gamma $) can be achieved at a minimizer called the ground state solution of (\ref{eq-1.2}). It is obvious that $E\left[ u,v \right]\ge -\left| \Gamma  \right|$ for $u,v\in {{H}^{1}}\left( {{\mathbb{R}}^{2}} \right),\text{ }I\left( u,v \right)=1$, which implies the value ${{\mu }_{\Gamma }}\ge -\left| \Gamma  \right|>-\infty $. Note that for the NLS equations with power nonlinearity, the infimum energy may not exist for some power magnitudes (see \cite{lions}). Here, as $\Gamma <0$, i.e., the self-focusing case, the potential energy of the saturable nonlinearity \[\int_{{{\mathbb{R}}^{2}}}{\Gamma \left[ {{\rho }^{2}}-ln\left( 1+{{\rho }^{2}} \right) \right]}\] may compete with the kinetic energy \[\int_{{{\mathbb{R}}^{2}}}{{{\left| \nabla \rho  \right|}^{2}}},\] so the magnitude of $\Gamma $ affects the existence of the ground state solution.
The main results may be stated as follows:
\BT\label{thmI}
Let ${{T}_{0}}$ be the following positive constant:
\[{{T}_{0}}=\underset{\begin{matrix}
   w\in {{H}^{1}}\left( {{\mathbb{R}}^{2}} \right)  \\
   {{\left\| w \right\|}_{2}}=1  \\
\end{matrix}}{\mathop inf}\,\frac{\int_{{{\mathbb{R}}^{2}}}{{{\left| \nabla w \right|}^{2}}}}{\int_{{{\mathbb{R}}^{2}}}{\left[ {{w}^{2}}-ln\left( 1+{{w}^{2}} \right) \right]}}\]
\begin{enumerate}
\item[(i)]~~If $\Gamma >-{{T}_{0}}$, then ${{\mu }_{\Gamma }}$ can not be attained by a minimizer i.e. there is no ground state solution.
\item[(ii)]~If $\Gamma <-{{T}_{0}}$, then ${{\mu }_{\Gamma }}<0$ and there exists a ground state solution which is radially symmetric and is denoted by $\left( u,v \right)=\rho \left( r \right)\left( \cos\phi ,\sin\phi  \right)$, where $\phi \in \mathbb{R}$ is an arbitrary constant and $\rho =\rho \left( r \right)$ is the energy minimizer of the following problem:
\BE\label{eq-1.7}
\text{Minimize }H\left[ \rho  \right]\text{   over   }\rho \in {{H}^{1}}\left( {{\mathbb{R}}^{2}} \right),\text{ }\int_{{{\mathbb{R}}^{2}}}{{{\rho }^{2}}}=1,
\EE
where
\BE\label{eq-1.8}
H\left[ \rho  \right]=\int_{{{\mathbb{R}}^{2}}}{{{\left| \nabla \rho  \right|}^{2}}+\Gamma \left[ {{\rho }^{2}}-ln\left( 1+{{\rho }^{2}} \right) \right].}
\EE
\end{enumerate}
\ET

The physical meaning is as follows: Theorem \ref{thmI} indicates that ground states only can behave like bright solitary waves in saturable photorefractive media. The constant $-{{T}_{0}}$ is the threshold for the existence of ground state solutions, which may be changed under the effect of external intensity ${{I}_{b}}$ (see Theorem \ref{thmI}~(ii)).

\BR\label{rk1}
By Schwartz symmetrization, it is obvious that the minimizer $\rho $ of the problem (\ref{eq-1.4}) must be radially symmetric and its Euler-Lagrange equation may be expressed as follows:
\BE\label{eq-1.9}
\left\{ \begin{matrix}
   {\rho }''+\frac{1}{r}{\rho }'-\Gamma \frac{{{\rho }^{3}}}{1+{{\rho }^{2}}}=\lambda \rho \text{   for  }r>0,  \\
   {\rho }'\left( 0 \right)=0,\text{  }\rho \left( 0 \right)>0,  \\
\end{matrix} \right.
\EE
where $\lambda $ is the Lagrange multiplier of the problem (\ref{eq-1.4}). We shall show that $\lambda >0\text{  }$ and the minimizer $\rho $ is a positive and monotone decreasing function which decays to zero exponentially as the variable $r$ goes to infinity.
\ER

\section{Proof of Theorem~\ref{thmI}}\label{s3}
\medskip
\noindent The proof is divided in a number of steps. The following lemma is crucial in proving Theorem \ref{thmI}.
\BL\label{LemmaI}
The value ${{\mu }_{\Gamma }}$ defined in (\ref{eq-1.6}) satisfies
\[{{\mu }_{\Gamma }}=\text{inf}\left\{ H\left[ \rho  \right]:\text{ }\rho \in {{H}^{1}}\left( {{\mathbb{R}}^{2}} \right),\text{ }\int_{{{\mathbb{R}}^{2}}}{{{\rho }^{2}}}=1 \right\}.\text{       }\]
\EL
\begin{proof}
Let $u=\rho \cos\phi \text{ and }v=\rho \sin\phi $, where both $\rho \text{ and }\phi $ are ${{H}^{1}}$ functions. Then \[{{\left| \nabla u \right|}^{2}}+{{\left| \nabla v \right|}^{2}}={{\left| \nabla \rho  \right|}^{2}}+{{\rho }^{2}}{{\left| \nabla \phi  \right|}^{2}}\] and ${{u}^{2}}+{{v}^{2}}={{\rho }^{2}}$, and hence the proof of Lemma \ref{LemmaI} is obvious.
\end{proof}

\subsection{Proof of Theorem~\ref{thmI}~(i)}\label{s3-1}
\BP\label{prop-a}
Let \[{{T}_{0}}=\underset{\begin{matrix}
   w\in {{H}^{1}}\left( {{\mathbb{R}}^{2}} \right)  \\
   {{\left\| w \right\|}_{2}}=1  \\
\end{matrix}}{\mathop inf}\,\frac{\int_{{{\mathbb{R}}^{2}}}{{{\left| \nabla w \right|}^{2}}}}{\int_{{{\mathbb{R}}^{2}}}{\left[ {{w}^{2}}-ln\left( 1+{{w}^{2}} \right) \right]}}.\] Then $T_0>0$.
\EP
\noindent To prove Proposition~\ref{prop-a}, we need

\noindent {\bf Claim~A1.}~~$\underset{s>0}{\mathop{\sup }}\,\frac{s-ln\left( 1+s \right)}{{{s}^{2}}}=\frac{1}{2}$.
\begin{proof}
Let $h\left( s \right)=\frac{s-ln\left( 1+s \right)}{{{s}^{2}}}$ for $s>0$. Then ${h}'\left( s \right)=-\frac{2+s}{{{s}^{2}}\left( 1+s \right)}+\frac{2}{{{s}^{3}}}ln\left( 1+s \right)$ and ${{\left( {{s}^{3}}{h}'\left( s \right) \right)}^{\prime }}=-\frac{{{s}^{2}}}{{{\left( 1+s \right)}^{2}}}<0$ for $s>0$, which implies ${h}'\left( s \right)<0$ for $s>0$, i.e. $h$ is a monotone decreasing function for $s>0$. Note that ${{s}^{3}}{h}'\left( s \right)=0\text{  at  }s=0$. On the other hand, by direct calculation, $\underset{s\to 0+}{\mathop{lim}}\,h\left( s \right)=\frac{1}{2}$ and hence the proof is complete .
\end{proof}
By Claim A1, ${{w}^{2}}-ln\left( 1+{{w}^{2}} \right)\le \frac{1}{2}{{w}^{4}}$, so $\int_{{{\mathbb{R}}^{2}}}{\left[ {{w}^{2}}-ln\left( 1+{{w}^{2}} \right) \right]}\le \frac{1}{2}\int_{{{\mathbb{R}}^{2}}}{{{w}^{4}}}$ for $w\in {{H}^{1}}\left( {{\mathbb{R}}^{2}} \right),\text{ }{{\left\| w \right\|}_{2}}=1$,
which implies
\[{{T}_{0}}\ge \underset{\begin{matrix}
   w\in {{H}^{1}}\left( {{\mathbb{R}}^{2}} \right)  \\
   {{\left\| w \right\|}_{2}}=1  \\
\end{matrix}}{\mathop{inf}}\,\frac{\int_{{{\mathbb{R}}^{2}}}{{{\left| \nabla w \right|}^{2}}}}{\frac{1}{2}\left\| w \right\|_{4}^{4}}\,.\]
On the other hand, by (2.2.5) in~\cite{LS-bk10},
\[\underset{\begin{matrix}
   w\in {{H}^{1}}\left( {{\mathbb{R}}^{2}} \right)  \\
   {{\left\| w \right\|}_{2}}=1  \\
\end{matrix}}{\mathop{inf}}\,\frac{\int_{{{\mathbb{R}}^{2}}}{{{\left| \nabla w \right|}^{2}}}}{\frac{1}{2}\left\| w \right\|_{4}^{4}}=2{{S}_{2,4}}>0\,,\]
where $S_{2,4}$ is the Sobolev constant. Therefore, $T_0\geq 2S_{2,4}>0$ and we complete the proof of Proposition~\ref{prop-a}.

\BP\label{prop-a2}
Suppose $\Gamma \in \left( -{{T}_{0}},0 \right)$ i.e. $0>\Gamma >-{{T}_{0}}$. Then the value ${{\mu }_{\Gamma }}$ can not attain a minimizer such that ${{\mu }_{\Gamma }}\le 0$.
\EP
\begin{proof}
We may prove by contradiction. Suppose there exists $u$ a minimizer of the value ${{\mu }_{\Gamma }}$ such that ${{\mu }_{\Gamma }}\le 0$. Then $\int_{{{\mathbb{R}}^{d}}}{{{\left| \nabla u \right|}^{2}}}+\Gamma \int_{{{\mathbb{R}}^{d}}}{\left[ {{u}^{2}}-ln\left( 1+{{u}^{2}} \right) \right]}\le 0$ and ${{\left\| u \right\|}_{2}}=1$. Hence
\[\int_{{{\mathbb{R}}^{d}}}{{{\left| \nabla u \right|}^{2}}}\le -\Gamma \int_{{{\mathbb{R}}^{d}}}{\left[ {{u}^{2}}-ln\left( 1+{{u}^{2}} \right) \right]}=\frac{-\Gamma }{{{T}_{0}}}\left\{ {{T}_{0}}\int_{{{\mathbb{R}}^{d}}}{\left[ {{u}^{2}}-ln\left( 1+{{u}^{2}} \right) \right]} \right\}\le \frac{-\Gamma }{{{T}_{0}}}\int_{{{\mathbb{R}}^{d}}}{{{\left| \nabla u \right|}^{2}}},\] which implies $u\equiv 0$, since $0<\frac{-\Gamma }{{{T}_{0}}}<1$. However, $u\equiv 0$ contradicts ${{\left\| u \right\|}_{2}}=1$. Therefore, we have completed the proof.
\end{proof}

\BP\label{prop-a3}
Suppose $\Gamma \in \left( -{{T}_{0}},0 \right)$, i.e. $0>\Gamma >-{{T}_{0}}$. Then the value ${{\mu }_{\Gamma }}$ can not attain a minimizer such that ${{\mu }_{\Gamma }}>0$.
\EP
\begin{proof}
It can be proved by contradiction. Suppose there exists a minimizer $u$ of the value ${{\mu }_{\Gamma }}$ such that ${{\mu }_{\Gamma }}>0$. Then $u$ satisfies the Euler-Lagrange equation of the problem \[{{\mu }_{\Gamma }}=\text{inf}\left\{ H\left[ \rho  \right]:\text{ }\rho \in {{H}^{1}}\left( {{\mathbb{R}}^{2}} \right),\text{ }\int_{{{\mathbb{R}}^{2}}}{{{\rho }^{2}}}=1 \right\}\] given by
\BE\label{eq-u}
\Delta u-\Gamma \frac{{{u}^{3}}}{1+{{u}^{2}}}=\lambda u\text{   for  }x\in {{\mathbb{R}}^{2}}
\EE
with ${{\left\| u \right\|}_{2}}=1$ and $u\left( x \right)\to 0$ as $\left| x \right|\to \infty $, where $\lambda $ is the Lagrange multiplier. Multiplying equation (\ref{eq-u}) by $x\cdot \nabla u$ and integrating over ${{\mathbb{R}}^{2}}$, we may derive the Pohozaev identity as follows:
\BE\label{pz1}
\lambda =-\Gamma \int_{{{\mathbb{R}}^{2}}}{\left[ {{u}^{2}}-ln\left( 1+{{u}^{2}} \right) \right]}\,.
\EE
The derivation is quite standard, so we omit the details here. On the other hand, multiplying equation (\ref{eq-u}) by $u$ and integrating over ${{\mathbb{R}}^{2}}$, gives
\BE\label{pz2}
\lambda =-\int_{{{\mathbb{R}}^{2}}}{{{\left| \nabla u \right|}^{2}}-\Gamma \int_{{{\mathbb{R}}^{2}}}{\frac{{{u}^{4}}}{1+{{u}^{2}}}}}.
\EE
Here we have used integration by parts. Suppose ${{\mu }_{\Gamma }}>0$. Then it is obvious that
\[\Gamma \int_{{{\mathbb{R}}^{2}}}{\left[ {{u}^{2}}-ln\left( 1+{{u}^{2}} \right) \right]}>-\int_{{{\mathbb{R}}^{2}}}{{{\left| \nabla u \right|}^{2}}}\,.\]
Hence (\ref{pz2}) implies
\BE\label{pz2-1}
\lambda <\Gamma \int_{{{\mathbb{R}}^{2}}}{\left[ {{u}^{2}}-ln\left( 1+{{u}^{2}} \right) \right]}-\Gamma \int_{{{\mathbb{R}}^{2}}}{\frac{{{u}^{4}}}{1+{{u}^{2}}}}.
\EE
Combining with (\ref{pz1}) and (\ref{pz2-1}), we have
$$
-2\Gamma \int_{{{\mathbb{R}}^{2}}}{\left[ {{u}^{2}}-ln\left( 1+{{u}^{2}} \right) \right]}<-\Gamma \int_{{{\mathbb{R}}^{2}}}{\frac{{{u}^{4}}}{1+{{u}^{2}}}},
$$
which is equivalent to
\BE\label{pz2-2}
2\int_{{{\mathbb{R}}^{2}}}{\left[ {{u}^{2}}-ln\left( 1+{{u}^{2}} \right) \right]}<\int_{{{\mathbb{R}}^{2}}}{\frac{{{u}^{4}}}{1+{{u}^{2}}}},
\EE
since $\Gamma <0$. Let \[F\left( s \right)=2\left[ s-ln\left( 1+s \right) \right]-\frac{{{s}^{2}}}{1+s}\text{   for   }s\ge 0\,.\] Then $F(0)=0$ and
$$
{F}'\left( s \right)=\frac{{{s}^{2}}}{{{\left( 1+s \right)}^{2}}}>0\text{   for   }s>0\,,
$$
and hence $F\left( s \right)\ge 0\text{  for  }s\ge 0$, i.e. \[2\left[ s-ln\left( 1+s \right) \right]\ge \frac{{{s}^{2}}}{1+s}\text{   for   }s\ge 0\,.\] Therefore, replacing $s$ by $u^2$, $2\int_{{{\mathbb{R}}^{2}}}{\left[ {{u}^{2}}-ln\left( 1+{{u}^{2}} \right) \right]}\ge \int_{{{\mathbb{R}}^{2}}}{\frac{{{u}^{4}}}{1+{{u}^{2}}}}$, which contradicts (\ref{pz2-2}) and we have completed the proof.
\end{proof}

\BP\label{prop-a4}
Suppose $\Gamma \ge 0$. Then ${{\mu }_{\Gamma }}=0$ can not attain a minimizer.
\EP
\begin{proof}
Let $w\in {{H}^{1}}\left( {{\mathbb{R}}^{2}} \right)\text{ and }{{\left\| w \right\|}_{2}}=1$. For $\delta >0$, let ${{w}_{\delta }}\left( x \right)=\delta w\left( \delta x \right)\text{  for  }x\in {{\mathbb{R}}^{2}}$. Then ${{\left\| {{w}_{\delta }} \right\|}_{2}}={{\left\| w \right\|}_{2}}=1$ and $\int_{{{\mathbb{R}}^{2}}}{{{\left| \nabla {{w}_{\delta }}\left( x \right) \right|}^{2}}}={{\delta }^{2}}\int_{{{\mathbb{R}}^{2}}}{{{\left| \nabla w\left( y \right) \right|}^{2}}dy}$. Moreover, by Claim A1, $\int_{{{\mathbb{R}}^{2}}}{\left[ w_{\delta }^{2}-ln\left( 1+w_{\delta }^{2} \right) \right]}\le \frac{1}{2}\int_{{{\mathbb{R}}^{2}}}{w_{\delta }^{4}}=\frac{1}{2}{{\delta }^{2}}\int_{{{\mathbb{R}}^{2}}}{{{w}^{4}}\left( y \right)dy}$. Hence $H\left[ {{w}_{\delta }} \right]=O\left( {{\delta }^{2}} \right)$ tends to zero as $\delta$ goes to zero. This implies that ${{\mu }_{\Gamma }}=0$. On the other hand, since $\Gamma \ge 0$, it is obvious that ${{\mu }_{\Gamma }}=0$ can not attain a minimizer. Therefore, the proof is completed.
\end{proof}
\noindent Combining Propositions~\ref{prop-a}-\ref{prop-a4}, completes the proof of Theorem~\ref{thmI}~(i).

\subsection{Proof of Theorem~\ref{thmI}~(ii)}\label{s3-2}
\medskip
\noindent

For the proof of Theorem~\ref{thmI}~(ii), we firstly consider the following problem:
$$
{{\mu }_{\Gamma ,\varepsilon }}=\underset{\begin{matrix}
   u\in H_{0}^{1}\left( {{B}_{\frac{1}{\varepsilon }}} \right)  \\
   {{P}_{\varepsilon }}\left[ u \right]=1  \\
\end{matrix}}{\mathop{inf}}\,{{H}_{\varepsilon }}\left[ u \right]\,,
$$
where \[{{H}_{\varepsilon }}\left[ u \right]=\int_{{{B}_{\frac{1}{\varepsilon }}}}{{{\left| \nabla u \right|}^{2}}+\Gamma \left[ {{u}^{2}}-ln\left( 1+{{u}^{2}} \right) \right]}\] and ${{P}_{\varepsilon }}\left[ u \right]=\int_{{{B}_{\frac{1}{\varepsilon }}}}{{{u}^{2}}}$ for $\varepsilon >0$ and $u\in H_{0}^{1}\left( {{B}_{\frac{1}{\varepsilon }}} \right)$. Hereafter, ${{B}_{\frac{1}{\varepsilon }}}$is the ball with radius $\frac{1}{\varepsilon }$ and center at the origin.

\BL\label{lemma2.1}
Assume $\Gamma <-{{T}_{0}}<0$. Then
\begin{enumerate}
\item[(i)]~~For $\varepsilon >0$, ${{\mu }_{\Gamma ,\varepsilon }}$ can be achieved by a minimizer ${{u}_{\varepsilon }}={{u}_{\varepsilon }}\left( r \right)>0$ with radial symmetry.
\item[(ii)]~For $\varepsilon >0$ sufficiently small, ${{\mu }_{\Gamma ,\varepsilon }}\le -{{c}_{0}}$, where ${{c}_{0}}$ is a positive constant independent of~$\varepsilon $.
\end{enumerate}
\EL
\begin{proof}
Fix $\varepsilon >0$ arbitrarily. Since $0\le {{s}^{2}}-ln\left( 1+{{s}^{2}} \right)\le C{{s}^{2}}$ for $s\in \mathbb{R}$, where $C$ is a positive constant independent of $s$, then ${{H}_{\varepsilon }}\left[ u \right]\ge \Gamma C\int_{{{B}_{\tfrac{1}{\varepsilon }}}}{{{u}^{2}}}=\Gamma C{{P}_{\varepsilon }}\left[ u \right]=\Gamma C$ for $u\in H_{0}^{1}\left( {{B}_{\tfrac{1}{\varepsilon }}} \right)$ and ${{P}_{\varepsilon }}\left[ u \right]=1$. Hence the value ${{\mu }_{\Gamma ,\varepsilon }}=\underset{\begin{matrix}
   u\in H_{0}^{1}\left( {{B}_{\tfrac{1}{\varepsilon }}} \right)  \\
   {{P}_{\varepsilon }}\left[ u \right]=1  \\
\end{matrix}}{\mathop{inf}}\,{{H}_{\varepsilon }}\left[ u \right]$ exists. Let $\left\{ {{v}_{k}} \right\}_{k=1}^{\infty }$ be a minimizing sequence of the value ${{\mu }_{\Gamma ,\varepsilon }}$. Without loss of generality, we may assume ${{H}_{\varepsilon }}\left[ {{v}_{k}} \right]\downarrow {{\mu }_{\Gamma ,\varepsilon }}$ as $k\rightarrow \infty $. Note that each ${{v}_{k}}\in H_{0}^{1}\left( {{B}_{\tfrac{1}{\varepsilon }}} \right)$ and ${{P}_{\varepsilon }}\left[ {{v}_{k}} \right]=1$. Apply symmetry rearrangement on each ${{v}_{k}}$, say $v_{k}^{*}\in H_{0}^{1}\left( {{B}_{\tfrac{1}{\varepsilon }}} \right)$ and ${{P}_{\varepsilon }}\left[ v_{k}^{*} \right]={{P}_{\varepsilon }}\left[ {{v}_{k}} \right]=1$. Note that $\int_{{{B}_{\tfrac{1}{\varepsilon }}}}{{{\left| \nabla v_{k}^{*} \right|}^{2}}}\le \int_{{{B}_{\tfrac{1}{\varepsilon }}}}{{{\left| \nabla {{v}_{k}} \right|}^{2}}}$ and $\int_{{{B}_{\tfrac{1}{\varepsilon }}}}{\left[ {{\left( v_{k}^{*} \right)}^{2}}-ln\left( 1+{{\left( v_{k}^{*} \right)}^{2}} \right) \right]}=\int_{{{B}_{\tfrac{1}{\varepsilon }}}}{\left[ {{\left( {{v}_{k}} \right)}^{2}}-ln\left( 1+{{\left( {{v}_{k}} \right)}^{2}} \right) \right]}$. Then ${{\mu }_{\Gamma ,\varepsilon }}\le {{H}_{\varepsilon }}\left[ v_{k}^{*} \right]\le {{H}_{\varepsilon }}\left[ {{v}_{k}} \right]$ for all $k$. Hence we may replace ${{v}_{k}}$ by $v_{k}^{*}$ and regard ${{v}_{k}}$'s as functions with radial symmetry. On the other hand, since  ${{H}_{\varepsilon }}\left[ {{v}_{k}} \right]\downarrow {{\mu }_{\Gamma ,\varepsilon }}$ as $k\rightarrow \infty $ and $0\le {{s}^{2}}-ln\left( 1+{{s}^{2}} \right)\le C{{s}^{2}}$ for $s\in \mathbb{R}$, then \[\int_{{{B}_{\tfrac{1}{\varepsilon }}}}{{{\left| \nabla {{v}_{k}} \right|}^{2}}}\le {{C}_{1}},\] where $C_1$ is a positive constant independent of $k$. Hence Sobolev embedding gives ${{v}_{k}}\to {{u}_{\varepsilon }}$ weakly in $H_{0}^{1}\left( {{B}_{\tfrac{1}{\varepsilon }}} \right)$ as $k\rightarrow \infty $ (up to a subsequence). Moreover, by the Sobolev compact embedding, ${{v}_{k}}\to {{u}_{\varepsilon }}$ in ${{L}^{2}}\left( {{B}_{\tfrac{1}{\varepsilon }}} \right)$ as $k\rightarrow \infty $. Consequently, ${{P}_{\varepsilon }}\left[ {{u}_{\varepsilon }} \right]=\underset{k\to \infty }{\mathop{\lim }}\,{{P}_{\varepsilon }}\left[ {{v}_{k}} \right]=1$ and ${{\mu }_{\Gamma ,\varepsilon }}\le {{H}_{\varepsilon }}\left[ {{u}_{\varepsilon }} \right]\le \underset{k\to \infty }{\mathop{\lim \inf }}\,{{H}_{\varepsilon }}\left[ {{v}_{k}} \right]={{\mu }_{\Gamma ,\varepsilon }}$. This implies that ${{u}_{\varepsilon }}$ is the minimizer of the value ${{\mu }_{\Gamma ,\varepsilon }}$. Here, we have used Fatou's Lemma. Moreover, since each $v_k$ is radially symmetric and ${{v}_{k}}\to {{u}_{\varepsilon }}$ in ${{L}^{2}}\left( {{B}_{\tfrac{1}{\varepsilon }}} \right)$ as $k\rightarrow \infty $, then ${{u}_{\varepsilon }}$ is radially symmetric. This completes the proof of Lemma~\ref{lemma2.1}~(i).

The proof of Lemma~\ref{lemma2.1}~(ii) is given as follows: From the definition of $T_0$,
$$
\forall \delta >0,\exists {{w}_{\delta }}\in {{H}^{1}}\left( {{\mathbb{R}}^{2}} \right),{{\left\| {{w}_{\delta }} \right\|}_{2}}=1\text{ such that }{{T}_{0}}>\frac{\int_{{{\mathbb{R}}^{2}}}{{{\left| \nabla {{w}_{\delta }} \right|}^{2}}}}{\int_{{{\mathbb{R}}^{2}}}{w_{\delta }^{2}-ln\left( 1+w_{\delta }^{2} \right)}}-\delta \,.
$$
Let $\delta =-\frac{\Gamma +{{T}_{0}}}{2}$. Then $\delta >0$ and $\Gamma <-{{T}_{0}}-\delta $, i.e. $-\Gamma >{{T}_{0}}+\delta $ due to $\Gamma <-{{T}_{0}}$. Hence $-\Gamma >\frac{\int_{{{\mathbb{R}}^{2}}}{{{\left| \nabla {{w}_{\delta }} \right|}^{2}}}}{\int_{{{\mathbb{R}}^{2}}}{w_{\delta }^{2}-ln\left( 1+w_{\delta }^{2} \right)}}$, i.e. $H\left( {{w}_{\delta }} \right)<0$. Let ${{w}_{\delta ,\varepsilon }}=\frac{{{w}_{\delta }}{{\varphi }_{\varepsilon }}}{{{\left\| {{w}_{\delta }}{{\varphi }_{\varepsilon }} \right\|}_{2}}}$, where ${{\varphi }_{\varepsilon }}\in C_{0}^{\infty }\left( {{\mathbb{R}}^{2}} \right)$ is a cut-off function such that: ${{\varphi }_{\varepsilon }}=1\text{  in  }{{B}_{\tfrac{1}{\varepsilon }-1}}$, ${{\varphi }_{\varepsilon }}=0\text{  in  }B_{\tfrac{1}{\varepsilon }}^{c}={{\mathbb{R}}^{2}}-{{B}_{\tfrac{1}{\varepsilon }}}$ and ${{\left\| {{\varphi }_{\varepsilon }} \right\|}_{{{C}^{1}}}}=O\left( 1 \right)$, where $O\left( 1 \right)$ is a bounded quantity. Then ${{w}_{\delta ,\varepsilon }}\in H_{0}^{1}\left( {{B}_{\tfrac{1}{\varepsilon }}} \right)$ and ${{\left\| {{w}_{\delta ,\varepsilon }} \right\|}_{2}}=1$. Note that $\varepsilon \text{ and }\delta $ are independent of each other. It is obvious that $\int_{B_{\tfrac{1}{\varepsilon }-1}^{c}}{w_{\delta }^{2}+{{\left| \nabla {{w}_{\delta }} \right|}^{2}}}={{o}_{\varepsilon }}\left( 1 \right)$, where $B_{\tfrac{1}{\varepsilon }-1}^{c}={{\mathbb{R}}^{2}}-{{B}_{\tfrac{1}{\varepsilon }-1}}$ and ${{o}_{\varepsilon }}\left( 1 \right)$ is a small quantity tending to zero as $\varepsilon $ goes to zero. Moreover, ${{\left\| {{w}_{\delta }}{{\varphi }_{\varepsilon }} \right\|}_{2}}={{\left\| {{w}_{\delta }} \right\|}_{2}}+{{o}_{\varepsilon }}\left( 1 \right)=1+{{o}_{\varepsilon }}\left( 1 \right)$, ${{\int_{{{\mathbb{R}}^{2}}}{\left| \nabla \left( {{w}_{\delta }}{{\varphi }_{\varepsilon }} \right) \right|}}^{2}}={{\int_{{{\mathbb{R}}^{2}}}{\left| \nabla {{w}_{\delta }} \right|}}^{2}}\varphi _{\varepsilon }^{2}+\int_{{{\mathbb{R}}^{2}}}{2\left( {{w}_{\delta }}\nabla {{w}_{\delta }} \right)}\cdot \left( {{\varphi }_{\varepsilon }}\nabla {{\varphi }_{\varepsilon }} \right)+w_{\delta }^{2}{{\left| \nabla {{\varphi }_{\varepsilon }} \right|}^{2}}={{\int_{{{\mathbb{R}}^{2}}}{\left| \nabla {{w}_{\delta }} \right|}}^{2}}+{{o}_{\varepsilon }}\left( 1 \right)$, and $\int_{{{\mathbb{R}}^{2}}}{\left\{ {{\left( {{w}_{\delta }}{{\varphi }_{\varepsilon }} \right)}^{2}}-\ln \left[ 1+{{\left( {{w}_{\delta }}{{\varphi }_{\varepsilon }} \right)}^{2}} \right] \right\}}=\int_{{{\mathbb{R}}^{2}}}{\left[ w_{\delta }^{2}-ln\left( 1+w_{\delta }^{2} \right) \right]}+{{o}_{\varepsilon }}\left( 1 \right)$. Hence from $H\left[ {{w}_{\delta }} \right]<0$, we get ${{H}_{\varepsilon }}\left[ {{w}_{\delta ,\varepsilon }} \right]=H\left[ {{w}_{\delta }} \right]+{{o}_{\varepsilon }}\left( 1 \right)\le -{{c}_{0}}$ as $\varepsilon $ gets sufficiently small, where ${{c}_{0}}=-\frac{1}{2}H\left[ {{w}_{\delta }} \right]>0$ is a constant independent of $\varepsilon $. Therefore, ${{\mu }_{\Gamma ,\varepsilon }}\le {{H}_{\varepsilon }}\left[ {{w}_{\delta ,\varepsilon }} \right]\le -{{c}_{0}}$ and we complete the proof of Lemma~\ref{lemma2.1}~(ii).
\end{proof}

\BR\label{remk}
From Lemma~\ref{lemma2.1}~(ii), we get ${{\mu }_{\Gamma }}\le {{\mu }_{\Gamma ,\varepsilon }}\le -{{c}_{0}}<0$.
\ER

\BL\label{lemma2.2}
Under the same hypothesis of Lemma~\ref{lemma2.1}, the minimizer ${{u}_{\varepsilon }}$ satisfies ${{\left\| {{u}_{\varepsilon }} \right\|}_{{{H}^{1}}\left( {{B}_{\tfrac{1}{\varepsilon }}} \right)}}\le {{K}_{0}}$ for $\varepsilon >0$ sufficiently small, where $K_0$ is a positive constant independent of $\varepsilon $.
\EL
\begin{proof}
By Lemma~\ref{lemma2.1}~(ii) ${{\mu }_{\Gamma ,\varepsilon }}\le -{{c}_{0}}$, which implies
$$
-{{c}_{0}}\ge {{\int_{{{B}_{\tfrac{1}{\varepsilon }}}}{\left| \nabla {{u}_{\varepsilon }} \right|}}^{2}}+\Gamma \int_{{{B}_{\tfrac{1}{\varepsilon }}}}{u_{\varepsilon }^{2}-ln\left( 1+u_{\varepsilon }^{2} \right)}\ge {{\int_{{{B}_{\tfrac{1}{\varepsilon }}}}{\left| \nabla {{u}_{\varepsilon }} \right|}}^{2}}+\Gamma\,.
$$
Here we have used the fact that $\int_{{{B}_{\tfrac{1}{\varepsilon }}}}{u_{\varepsilon }^{2}}=1$. Thus
$$\left\| {{u}_{\varepsilon }} \right\|_{{{H}^{1}}\left( {{B}_{\tfrac{1}{\varepsilon }}} \right)}^{2}={{\int_{{{B}_{\tfrac{1}{\varepsilon }}}}{\left| \nabla {{u}_{\varepsilon }} \right|}}^{2}}+\int_{{{B}_{\tfrac{1}{\varepsilon }}}}{u_{\varepsilon }^{2}}\le 1-{{c}_{0}}-\Gamma
$$ and we have completed the proof.
\end{proof}

We may extend ${{u}_{\varepsilon }}$ on the entire plane ${{\mathbb{R}}^{2}}$ by setting ${{u}_{\varepsilon }}\left( r \right)=0\text{  for  }r>\tfrac{1}{\varepsilon }$. Note that each ${{u}_{\varepsilon }}$ is radially symmetric. Then Lemma~\ref{lemma2.2} gives ${{\left\| {{u}_{\varepsilon }} \right\|}_{H_{r}^{1}\left( {{\mathbb{R}}^{2}} \right)}}\le {{K}_{0}}$, which implies
\BE\label{conv1}
{{u}_{\varepsilon }}\to U\text{  weakly in   }H_{r}^{1}\left( {{\mathbb{R}}^{2}} \right)
\EE
as $\varepsilon $ goes to zero (up to a subsequence).
Furthermore, by the compact embedding of $H^1_r$ radial functions to $L^4_r$ functions (cf. Lions paper \cite{lions}), we have
\BE\label{conv2}
{{u}_{\varepsilon }}\to U\text{  in   }L_{r}^{4}\left( {{\mathbb{R}}^{2}} \right)
\EE
as $\varepsilon $ goes to zero (up to a subsequence).
Now, we want to prove that $U$ is nontrivial. By Claim~A1, we get
$u_{\varepsilon }^{2}-ln\left( 1+u_{\varepsilon }^{2} \right)\le \tfrac{1}{2}u_{\varepsilon }^{4}$. Hence by (\ref{conv1}), (\ref{conv2}) and Lemma~\ref{lemma2.1}~(ii),
\begin{align}
& -{{c}_{0}}\ge \underset{\varepsilon \to 0+}{\mathop{liminf}}\,{{\mu }_{\Gamma ,\varepsilon }}=\underset{\varepsilon \to 0+}{\mathop{liminf}}\,{{\int_{{{\mathbb{R}}^{2}}}{\left| \nabla {{u}_{\varepsilon }} \right|}}^{2}}+\Gamma \left[ u_{\varepsilon }^{2}-ln\left( 1+u_{\varepsilon }^{2} \right) \right], \\
& \text{     }\ge \underset{\varepsilon \to 0+}{\mathop{liminf}}\,{{\int_{{{\mathbb{R}}^{2}}}{\left| \nabla {{u}_{\varepsilon }} \right|}}^{2}}+\tfrac{1}{2}\Gamma u_{\varepsilon }^{4}\ge {{\int_{{{\mathbb{R}}^{2}}}{\left| \nabla U \right|}}^{2}}+\tfrac{1}{2}\Gamma {{U}^{4}},
\end{align}
which implies that $U$ is nontrivial. Here we have used again Fatou's Lemma. Otherwise, $U\equiv 0$ and then $0>-c_0\geq 0$, a contradiction.

To complete the proof, we have to prove that ${{\left\| U \right\|}_{2}}=1$. The idea is to use the concentration compactness. We shall prove that the class ${{\left\{ {{u}_{\varepsilon }} \right\}}_{\varepsilon >0}}$ is neither vanishing nor dichotomy as $\varepsilon \to 0$ by contradiction. Suppose ${{\left\{ {{u}_{\varepsilon }} \right\}}_{\varepsilon >0}}$ is vanishing i.e. ${{u}_{\varepsilon }}\to 0\text{  in  }{{L}^{2}}\left( {{\mathbb{R}}^{2}} \right)$ as $\varepsilon \to 0$. Then by Egoroff Theorem, ${{u}_{\varepsilon }}\to 0\text{  almost everywhere}$ as $\varepsilon \to 0$ (up to a subsequence). On the other hand, by (\ref{conv2}) and Egoroff Theorem, ${{u}_{\varepsilon }}\to U\text{  almost everywhere}$ as $\varepsilon \to 0$ (up to a subsequence). Consequently, $U\equiv 0$ but $U$ is nontrivial. This gives a contradiction, so the class ${{\left\{ {{u}_{\varepsilon }} \right\}}_{\varepsilon >0}}$ is not vanishing.

To show that ${{\left\{ {{u}_{\varepsilon }} \right\}}_{\varepsilon >0}}$ is not dichotomy as $\varepsilon \to 0$, we study the solution profile of ${{u}_{\varepsilon }}$. The minimizer ${{u}_{\varepsilon }}$ satisfies the following equation:
\BE\label{eq-mizr1}
\Delta {{u}_{\varepsilon }}-\Gamma \frac{u_{\varepsilon }^{3}}{1+u_{\varepsilon }^{2}}={{\lambda }_{\varepsilon }}{{u}_{\varepsilon }}\text{   in  }{{B}_{\tfrac{1}{\varepsilon }}}
\EE
with the zero Dirichlet boundary condition ${{u}_{\varepsilon }}=0\text{  on  }\partial {{B}_{\tfrac{1}{\varepsilon }}}$, where ${{\lambda }_{\varepsilon }}$ is the associated Lagrange multiplier. Multiply equation (\ref{eq-mizr1}) by ${{u}_{\varepsilon }}$ and integrate over ${{B}_{\tfrac{1}{\varepsilon }}}$. Then using integration by parts and $\int_{{{B}_{\tfrac{1}{\varepsilon }}}}{u_{\varepsilon }^{2}}=1$, we get ${{\lambda }_{\varepsilon }}={{\lambda }_{\varepsilon }}\int_{{{B}_{\tfrac{1}{\varepsilon }}}}{u_{\varepsilon }^{2}}=-{{\int_{{{B}_{\tfrac{1}{\varepsilon }}}}{\left| \nabla {{u}_{\varepsilon }} \right|}}^{2}}-\Gamma \int_{{{B}_{\tfrac{1}{\varepsilon }}}}{\frac{u_{\varepsilon }^{4}}{1+u_{\varepsilon }^{2}}}$. Hence by Lemma~\ref{lemma2.2},
\BE\label{est-ev}
\left| {{\lambda }_{\varepsilon }} \right|\le {{K}_{1}},
\EE
where $K_1$ is a positive constant independent of $\varepsilon $.

Since ${{u}_{\varepsilon }}={{u}_{\varepsilon }}\left( r \right)$ is radically symmetric, equation (\ref{eq-mizr1}) and the zero Dirichlet boundary condition can be reduced to a boundary value problem of an ordinary differential equation as follows:
\BE\label{eq-mizr2}
\left\{ \begin{matrix}
   {{{{u}''}}_{\varepsilon }}+\frac{1}{r}{{{{u}'}}_{\varepsilon }}-\Gamma \frac{u_{\varepsilon }^{3}}{1+u_{\varepsilon }^{2}}={{\lambda }_{\varepsilon }}{{u}_{\varepsilon }}\text{  for  }0<r<\tfrac{1}{\varepsilon },  \\
   {{{{u}'}}_{\varepsilon }}\left( 0 \right)=0,\text{                    }{{u}_{\varepsilon }}\left( \tfrac{1}{\varepsilon } \right)=0.  \\
\end{matrix} \right.
\EE
From the energy comparison, we may set ${{u}_{\varepsilon }}\left( r \right)\ge 0$ for $0<r<\tfrac{1}{\varepsilon }$.
\BL\label{lemma2.3}
The minimizer ${{u}_{\varepsilon }}={{u}_{\varepsilon }}\left( r \right)$ is positive and monotone decreasing with $r$.
\EL
\begin{proof}
   Suppose ${{\lambda }_{\varepsilon }}>0$. Then equation (\ref{eq-mizr1}) implies
$$
\Delta {{u}_{\varepsilon }}-{{\lambda }_{\varepsilon }}{{u}_{\varepsilon }}=\Gamma \frac{u_{\varepsilon }^{3}}{1+u_{\varepsilon }^{2}}\le 0\text{   in   }{{B}_{\tfrac{1}{\varepsilon }}}
$$
with the zero Dirichlet boundary condition ${{u}_{\varepsilon }}=0\text{  on  }\partial {{B}_{\tfrac{1}{\varepsilon }}}$. Hence by the strong maximum principle, ${{u}_{\varepsilon }}>0\text{   in  }{{B}_{\tfrac{1}{\varepsilon }}}$ and
$$
{{u}_{\varepsilon }}\left( r \right)=\underset{x\in \partial {{B}_{r}}}{\mathop{min}}\,{{u}_{\varepsilon }}\left( x \right)=\underset{x\in {{B}_{r}}}{\mathop{\min }}\,{{u}_{\varepsilon }}\left( x \right)>\underset{x\in {{B}_{s}}}{\mathop{\min }}\,{{u}_{\varepsilon }}\left( x \right)=\underset{x\in \partial {{B}_{s}}}{\mathop{min}}\,{{u}_{\varepsilon }}\left( x \right)={{u}_{\varepsilon }}\left( s \right)
$$
for $0<r<s<\tfrac{1}{\varepsilon }$ since ${{u}_{\varepsilon }}={{u}_{\varepsilon }}\left( r \right)$ is radically symmetric. This shows that ${{u}_{\varepsilon }}={{u}_{\varepsilon }}\left( r \right)$ is positive and monotone decreasing with $r$. Similarly, if ${{\lambda }_{\varepsilon }}\le 0$, then equation (\ref{eq-mizr1}) implies
$$
\Delta {{u}_{\varepsilon }}=\left( {{\lambda }_{\varepsilon }}+\Gamma \frac{u_{\varepsilon }^{2}}{1+u_{\varepsilon }^{2}} \right){{u}_{\varepsilon }}\le 0\text{   in   }{{B}_{\tfrac{1}{\varepsilon }}}
$$
with the zero Dirichlet boundary condition ${{u}_{\varepsilon }}=0\text{  on  }\partial {{B}_{\tfrac{1}{\varepsilon }}}$. Thus using the strong maximum principle again, we may prove that ${{u}_{\varepsilon }}={{u}_{\varepsilon }}\left( r \right)$ is positive and monotone decreasing with $r$, and thus complete the proof of Lemma~\ref{lemma2.3}.
\end{proof}
From Lemma~\ref{lemma2.3}, ${{u}_{\varepsilon }}$ can not be splitted into two parts as $\varepsilon \to 0+$, and hence ${{\left\{ {{u}_{\varepsilon }} \right\}}_{\varepsilon >0}}$ can not be dichotomy as $\varepsilon \to 0+$. Therefore, by the concentration compactness Theorem, ${{u}_{\varepsilon }}\to U\text{  in  }{{L}^{2}}\left( {{\mathbb{R}}^{2}} \right)$ as $\varepsilon \to 0+$ (up to a subsequence), which implies ${{\left\| U \right\|}_{2}}=1$ since ${{\left\| {{u}_{\varepsilon }} \right\|}_{2}}=1$.
Now we claim that the limit function $U$ satisfies
\BE\label{eq-U1}
\Delta U-\Gamma \frac{{{U}^{3}}}{1+{{U}^{2}}}={{\lambda }_{0}}U\text{   in   }{{\mathbb{R}}^{2}},
\EE
$U=U\left( r \right)\ge 0$ is radially symmetric, and $\underset{r\to \infty }{\mathop{\lim }}\,U\left( r \right)=0$, where ${{\lambda }_{0}}$ is the limit of ${{\lambda }_{\varepsilon }}$'s (up to a subsequence) since (\ref{est-ev}) implies
\BE\label{conv-ev}
{{\lambda }_{\varepsilon }}\to {{\lambda }_{0}}\text{   as   }\varepsilon \to \text{0+  (up to a subsequence)}.
\EE
Let $\phi \in C_{0}^{\infty }\left( {{\mathbb{R}}^{2}} \right)$ be any test function. Since ${{u}_{\varepsilon }}$ satisfies (\ref{eq-mizr1}), then
\BE\label{weq-minzr1}
\int_{{{\mathbb{R}}^{2}}}{\nabla {{u}_{\varepsilon }}\cdot \nabla \phi +\Gamma \int_{{{\mathbb{R}}^{2}}}{\frac{u_{\varepsilon }^{3}}{1+u_{\varepsilon }^{2}}\phi =-{{\lambda }_{\varepsilon }}\int_{{{\mathbb{R}}^{2}}}{{{u}_{\varepsilon }}\phi }}}\,.
\EE
Hence (\ref{conv1}) and (\ref{conv2}) give
$$\int_{{{\mathbb{R}}^{2}}}{\nabla {{u}_{\varepsilon }}\cdot \nabla \phi \to \int_{{{\mathbb{R}}^{2}}}{\nabla U\cdot \nabla \phi }}\,, $$
$$\int_{{{\mathbb{R}}^{2}}}{{{u}_{\varepsilon }}\phi \to \int_{{{\mathbb{R}}^{2}}}{U\phi }}\,, $$
and
\begin{align} \nonumber
  & \int_{{{\mathbb{R}}^{2}}}{\frac{u_{\varepsilon }^{3}}{1+u_{\varepsilon }^{2}}\phi =}\int_{{{\mathbb{R}}^{2}}}{{{u}_{\varepsilon }}\phi -\int_{{{\mathbb{R}}^{2}}}{\frac{{{u}_{\varepsilon }}\phi }{1+u_{\varepsilon }^{2}}}} \\
 & \text{                 =}\int_{{{\mathbb{R}}^{2}}}{{{u}_{\varepsilon }}\phi }-\int_{{{\mathbb{R}}^{2}}}{\frac{\phi }{1+u_{\varepsilon }^{2}}}\left( {{u}_{\varepsilon }}-U \right)-\int_{{{\mathbb{R}}^{2}}}{\frac{\phi }{1+u_{\varepsilon }^{2}}}U \nonumber \\
 & \text{                 =}\int_{{{\mathbb{R}}^{2}}}{{{u}_{\varepsilon }}\phi }-\int_{{{\mathbb{R}}^{2}}}{\frac{\phi }{1+u_{\varepsilon }^{2}}}\left( {{u}_{\varepsilon }}-U \right) \nonumber \\
 & \text{                    }-\int_{{{\mathbb{R}}^{2}}}{\frac{\phi }{1+{{U}^{2}}}}U-\int_{{{\mathbb{R}}^{2}}}{\left( U-{{u}_{\varepsilon }} \right)}\frac{U+{{u}_{\varepsilon }}}{\left( 1+u_{\varepsilon }^{2} \right)\left( 1+{{U}^{2}} \right)}U\phi  \nonumber \\
 & \text{               }\to \int_{{{\mathbb{R}}^{2}}}{U\phi -}\int_{{{\mathbb{R}}^{2}}}{\frac{\phi }{1+{{U}^{2}}}}U=\int_{{{\mathbb{R}}^{2}}}{\frac{{{U}^{3}}}{1+{{U}^{2}}}}\phi. \nonumber
\end{align}
Note that ${{\left\| \frac{\phi }{1+u_{\varepsilon }^{2}} \right\|}_{\tfrac{4}{3}}}\le {{\left\| \phi  \right\|}_{\tfrac{4}{3}}}$ and ${{\left\| \frac{U+{{u}_{\varepsilon }}}{\left( 1+u_{\varepsilon }^{2} \right)\left( 1+{{U}^{2}} \right)}U\phi  \right\|}_{\tfrac{4}{3}}}\le {{\left\| U\phi  \right\|}_{\tfrac{4}{3}}}\le {{\left\| U \right\|}_{4}}{{\left\| \phi  \right\|}_{2}}$ since $$\left| \frac{U+{{u}_{\varepsilon }}}{\left( 1+u_{\varepsilon }^{2} \right)\left( 1+{{U}^{2}} \right)} \right|\le \left| \frac{U}{\left( 1+u_{\varepsilon }^{2} \right)\left( 1+{{U}^{2}} \right)} \right|+\left| \frac{{{u}_{\varepsilon }}}{\left( 1+u_{\varepsilon }^{2} \right)\left( 1+{{U}^{2}} \right)} \right|\le \frac{1}{2}+\frac{1}{2}=1\,.$$
Thus (\ref{weq-minzr1}) and (\ref{conv-ev}) imply
$$\int_{{{\mathbb{R}}^{2}}}{\nabla U\cdot \nabla \phi }+\Gamma \int_{{{\mathbb{R}}^{2}}}{\frac{{{U}^{3}}}{1+{{U}^{2}}}}\phi =-{{\lambda }_{0}}\int_{{{\mathbb{R}}^{2}}}{U\phi }$$
and then $U$ satisfies (\ref{eq-U1}) and $\underset{\left| x \right|\to \infty }{\mathop{\lim }}\,U\left( x \right)=0$. Moreover, $U=U\left( r \right)\ge 0$ is radially symmetric, since each ${{u}_{\varepsilon }}$ is positive and radially symmetric. Therefore, the equation for $U$ can be written as
$$\left\{ \begin{matrix}
   {U}''+\frac{1}{r}{U}'-\Gamma \frac{{{U}^{3}}}{1+{{U}^{2}}}={{\lambda }_{0}}U\text{   for   }r>0,  \\
   {U}'\left( 0 \right)=0,\text{              }U\left( \infty  \right)\text{=0                 } . \\
\end{matrix} \right.$$
By the uniqueness of ordinary differential equations, we may assume $U\left( 0 \right)>0$.

\BL\label{lemma2.4}
$U\left( r \right)>0\text{   for   }r\ge 0$.
\EL
\begin{proof}
We may prove by contradiction. Suppose there exists ${{r}_{0}}>0$ a minimum point of $U$ such that $U\left( {{r}_{0}} \right)=0$. Then ${U}'\left( {{r}_{0}} \right)=U\left( {{r}_{0}} \right)=0$. Hence by the uniqueness of ordinary differential equations, $U\equiv 0$ contradicts ${{\left\| U \right\|}_{2}}=1$ and we have completed the proof.
\end{proof}

Due to $\underset{r\to \infty }{\mathop{\lim }}\,U\left( r \right)=0$, there exists $R_1>0$ such that $0<U\left( r \right)\le 1\text{   for   }r\ge {{R}_{1}}$. By equation (\ref{eq-U1}), $\Delta U=\left( \Gamma \frac{{{U}^{2}}}{1+{{U}^{2}}}+{{\lambda }_{0}} \right)U\in {{L}^{4}}\left( {{B}_{{{R}_{1}}}} \right)$, since $\Gamma \frac{{{U}^{2}}}{1+{{U}^{2}}}+{{\lambda }_{0}}\in {{L}^{\infty }}$. Hence by the standard regularity theorem of Poisson equation, $U\in {{W}^{2,4}}\left( {{B}_{{{R}_{1}}}} \right)$ and then by the Sobolev embedding ${{W}^{2,4}}\left( {{B}_{{{R}_{1}}}} \right)\subset {{L}^{\infty }}\left( {{B}_{{{R}_{1}}}} \right)$, we obtain
\BL\label{lemma2.5}
$U\left( r \right)\le {{K}_{2}}\text{   for  }r\ge 0$, where $K_2$ is a positive constant.
\EL
Now we prove that ${{\lambda }_{0}}$ is positive by contradiction. Suppose ${{\lambda }_{0}}\le 0$. Then equation (\ref{eq-U1}) and Lemma~\ref{lemma2.4} imply $$\Delta U={{\lambda }_{0}}U+\Gamma \frac{{{U}^{3}}}{1+{{U}^{2}}}\le 0\text{   in  }{{\mathbb{R}}^{2}}\,.$$ Hence by Lemma~\ref{lemma2.5} and the Liouville Theorem, $U$ must be a constant function, i.e. $U\equiv 0$, which is impossible. Therefore, we conclude that
\BL\label{lemma2.6}
${{\lambda }_{0}}>0$.
\EL
Since $\Gamma <0$, the equation~(\ref{eq-U1}) becomes $$\Delta U-{{\lambda }_{0}}U=\Gamma \frac{{{U}^{3}}}{1+{{U}^{2}}}\le 0\text{   in   }{{\mathbb{R}}^{2}}\,.$$ Hence, as for the proof of Lemma~\ref{lemma2.3}, we may use the strong maximum principle to prove that
\BL\label{lemma 2.7}
$U=U\left( r \right)$ is monotone decreasing with $r$.
\EL
In this manner, we have completed the proof of Theorem~\ref{thmI}~(ii), and with this the complete proof of Theorem~\ref{thmI}.

\section{Conclusion}
In this paper the existence of ground states in the form of counterpropagating solitons
in a multidimensional nonlinear Schr\"odinger model of paraxial beam propagation in media with saturable
nonlinearity has been proven. From the proof, a threshold condition on the beam coupling constant for the existence of such fundamental
solitons followed.\\

This work has been supported by the Qatar
National Research Fund project NPRP 09-462-1-074, by the Texas Norman Hackman
Advanced Research Program Grant No. 010366-0149-2009 from the Texas Higher
Education Coordinating Board, and by the grant from the National Science Council of Republic of China.

\end{document}